\documentclass[floatfix,preprint,showkeys,citeautoscript,showpacs,amsmath,amssymb,epsf,aip,jcp]{revtex4-1}
\usepackage{dcolumn}
\usepackage{epsfig}
\usepackage{color}
\usepackage{graphicx}
\usepackage{dcolumn}
\usepackage{bm}

\begin{document}
\title{Fermi Orbital Derivatives in Self-Interaction Corrected Density Functional Theory: Applications to Closed Shell Atoms}

\author{Mark R. Pederson}
\email{mark.pederson@science.doe.gov}
\affiliation{Department of Chemistry, Johns Hopkins University, Baltimore MD, 21218}                        
\date{\today }
\begin{abstract}
A recent modification of the Perdew-Zunger self-interaction-correction (SIC) to the  density-functional formalism~\cite{PRP} 
has provided a framework for explicitly restoring unitary invariance to the expression for the total energy. 
The formalism depends upon construction of L{\"o}wdin orthonormalized~\cite{Low} Fermi-orbitals~\cite{FO1,FO3} which parametrically
depend on variational quasi-classical electronic positions. Derivatives of these quasi-classical electronic
positions, required for efficient minimization of the
self-interaction corrected energy, are derived and tested here on atoms. 
Total energies and ionization
energies in closed-shell singlet atoms, where correlation is less important, using the PW92 LDA functional~\cite{PW92}, are in good 
agreement with experiment and non-relativistic Quantum-Monte-Carlo (QMC) results albeit slightly too low.
                                                                                                                               
\end{abstract}
\maketitle

\section{Introduction}
This paper derives equations necessary for testing a new modification~\cite{PRP}
of self-interaction corrected density functional theory~\cite{PZ} and
applies it to the calculation of total energies and the highest-occupied-molecular-orbital (HOMO) 
eigenvalues in atoms. The results show significant improvement in
the HOMO-level alignments, relative to experimental ionization energies, and also show that the resulting self-interaction
corrected total energies are significantly improved over density-functional approximations that do not include the self-interaction
correction.

The use of localized orbital sets in quantum-mechanical calculations has a long history in both many-electron 
wavefunction methods and density-functional calculations. Such orbital sets are often of interest from the 
standpoint of chemical and physical interpretations of bonding or from the perspective of localized excitations.    
In density-functional-based methods~\cite{HK1,HK2} 
and single-determinantal Hartree-Fock methods, many sets of physically appealing localized orthornormal orbitals 
may be constructed from a unitary transformation on eigenstates of the  single-particle Hamiltonian or Fock matrix~\cite{RE63}.
Each of these orbital sets lead to the 
same total energy since the spin densities, spin-density matrices, and Slater determinant do not change under the action 
of a unitary transformation on the occupied orbital space.  However, in orbital-dependent formulations 
such as self-interaction corrected
density functional methods, the  orbital dependence appearing in the expression for the total energy leads to an 
expression that is not invariant to unitary transformations within the occupied-orbital space.  
As a result, past implementations of SIC have used a double-iteration process which proceeds by first optimizing the 
canonical orbitals and then finding the best unitary matrix that minimizes 
the SIC part of the 
energy (Refs. ~\onlinecite{HarJPB83,HeaPRB83,PedJCP84,PedJCP85,PedJCP88,temmerman,simon1,simon2,hoffmann,MaiDinh,Suraud,RLT,Kummel,Iceland,Dabo}).
Despite a large number of implementational differences in
Refs.~\onlinecite{HarJPB83,HeaPRB83,PedJCP84,PedJCP85,PedJCP88,temmerman,simon1,simon2,hoffmann,MaiDinh,Suraud,RLT,Kummel,Iceland,Dabo} the 
general perspective of these articles is that the best possible set of localized orbitals should be found for minimizing the SIC 
energy. For the remainder of this paper this is referred to generically as the standard perspective or method.
The auxiliary conditions 
required for minimization of the energy have been referred to as the 
localization equations~\cite{PedJCP84,PedJCP85,PedJCP88} 
or symmetry conditions~\cite{MaiDinh,Suraud} and a variety of methods for solving these equations has  been 
suggested and successfully employed.  
The latter references~\cite{Kummel,Iceland,Dabo} contains significant discussion about present and past approaches
to this with a good number of applications based on modern numerical methods and also discuss the need for consideration of
complex unitary transformations (See also Ref.~\onlinecite{RLT}). 

As an alternative to the localization-equation-based formulation, which is not explicitly invariant under unitary 
transformations of the occupied orbitals, Pederson, Ruzsinszky and Perdew have introduced a technique 
for constructing unitary matrices that explicitly depend upon the density matrix.  This technique allows one 
to construct N localized orbitals 
from N quasi-classical electronic positions by first creating a set of Fermi Orbitals and then using L{\"o}wdin's method of 
symmetric orthonormalization to find a unitarily equivalent set of localized orbitals.   
The resulting orthonormal set of functions, for a small enough number of electronic shells,  
can coincide with Wannier functions in solids, sp$^n$ hybrids in atoms, and 
Edmiston-Ruedenberg equivalent orbitals in molecules. Moreover, they  have physical appearances that are similar to  
most of the orbitals that have been used in past SIC applications.

In Section II a brief review of the implementation of self-interaction corrections within the Perdew-Zunger formulation and the 
now standard approach to solving the self-consistent equations is provided. Then the equations 
needed for reposing the SIC functional in terms of LOs which parametrically depend on Fermi orbital 
centroids are presented.  Within this prescription derivatives of the energy with respect to the 
Fermi Orbital centroids are developed.                           
In Sec III,  the method is applied to a set of closed-shell atoms with non-degenerate ground 
states for which the Hohenberg-Kohn theorem is strongest.
The primary focus is to demonstrate that the analytical expressions are correct  and to provide   
results on atom 
total energies and the highest occupied eigenvalues.  
In Sec IV, results are discussed and some tentative conclusions are offered.

\section{Energies and Derivatives within Fermi-Orbital SIC}
Given any approximation to the density functional for the sum of the Coulomb and exchange-correlation
energies denoted by $F^{approx}[\rho_\uparrow,\rho_\downarrow]$), the Perdew-Zunger  
self-interaction corrected expression for a spin-polarized system is  written according to:
\begin{eqnarray}
E^{SIC-DFT}=F^{approx}[\rho_\uparrow,\rho_\downarrow]-\Sigma_{i\sigma} F^{approx}[\rho_{i\sigma},0] \\
\rho_\sigma({\bf r}) 
= \Sigma_i |\phi_{i\sigma}({\bf r})|^2
= \Sigma_\alpha |\psi_{\alpha \sigma}({\bf r})|^2
\end{eqnarray}
with the $N_\sigma$ localized orbital densities given by $\rho_{i\sigma}(\bf r)= |\phi_{i\sigma}({\bf r})|^2$. In the 
above equation the localized orbitals $\{\phi_{i\sigma}\}$ are constructed from an $N_\sigma \times N_\sigma$ dimensional
unitary transformation on the so-called canonical-orbital set $\{\psi_{\alpha \sigma}\}$. 
The canonical orbital set coincides exactly with the
Kohn-Sham orbitals in the limit that the SIC vanishes. An alternative to the computationally tractable 
two-step procedures for variational minimization of the energy is to 
to directly solve for the the localized orbital set which must self-consistently satisfy:
\begin{eqnarray}
\{H_{o\sigma} + V^{SIC}_{i \sigma} \}|\phi_{i\sigma}> = \Sigma_j \lambda^\sigma_{ij}|\phi_{j\sigma}>  \\
<\phi_{i\sigma}| V^{SIC}_{i \sigma} - V^{SIC}_{j \sigma} |\phi_{j\sigma}> = 0. \label{loq}
\end{eqnarray}

However to avoid having to solve the localization equation (Eq.~\ref{loq}) and restore unitary invariance,
the localized orbitals may instead be  derived from the Fermi orbitals (FO)~\cite{FO1,FO3} which 
depend parametrically on a classical electronic position  but are explicitly determined from the density-matrix:
\begin{eqnarray}
F_{i}(\bf r)=\frac{\rho({\bf a}_{i},{\bf r}) }{\sqrt{\rho({\bf a}_{i})}}, \\
F_{i}({\bf r})=
\frac{\Sigma_{\alpha}  \psi^{*}_{\alpha}
({\bf a}_{i})
\psi_{\alpha}(\bf r)}{\sqrt{\{ \Sigma_{\alpha}|\psi_{\alpha} \bf(a_{i})|^2\}}}
\equiv\Sigma_{\alpha}F_{i\alpha}^{\psi}
\psi_{\alpha}({\bf r})\label{eqn2}.
\end{eqnarray}
In the above equation, the classical electronic positions $\{{\bf a}_{i}\}$ then become variational parameters for minimizing
the orbital-dependent part of the energy.
The density of the Fermi orbital is exactly equal to the spin density at its classical centroid and it is automatically
normalized since the Kohn-Sham orbitals are orthonormal. 
The approach suggested in Ref~\onlinecite{PRP} was to use
L{\"o}wdin's method of symmetric orthonormalization~\cite{Low}, to determine an  orbital set
$\{\phi_{1\sigma},\phi_{2\sigma}...,\phi_{N_\sigma \sigma}\}$ than can be used to 
construct the SIC-DFT energy of Eq.~(1).  

In Ref.~\onlinecite{PRP} minimization of the energy in molecules by brute force adjustment 
of the FO centroids lead to improved atomization energies. 
However, it was noted that with derivatives it would be possible to use methods 
such as conjugate gradients to variationally determine the FOC. In order to do this, one starts
by writing:
\begin{eqnarray}
\frac{dE^{SIC}}{d a_{m}}
=\Sigma_k \{ 
< \frac{d \phi_k}{ d a_{m} } | V_k^{SIC} |\phi_k> +  
<\phi_k  | V_k^{SIC} | \frac{d \phi_k}{ d a_{m} }> \}
\label{eqn9}\\
\frac{dE^{SIC}}{d a_{m}}
=\Sigma_{kl} \{ 
< \frac{d \phi_k}{ d a_{m} } |\phi_l><\phi_l| V_k^{SIC} | \phi_k> +  
<\phi_k | V_k^{SIC} | \phi_l><\phi_l|  \frac{d \phi_k}{ d a_{m} }> \}
\label{eqn9a}\\
\frac{dE^{SIC}}{d a_{m}}
=\Sigma_{kl}  
\epsilon^{k}_{kl}\{
< \frac{d \phi_k}{ d a_{m} } |\phi_l> +  <\phi_l|  \frac{d \phi_k}{ d a_{m} }> \}
\label{eqn9b}\\
\frac{dE^{SIC}}{d a_{m}}
=\Sigma_{kl}  
\epsilon^{k}_{kl}\{
< \frac{d \phi_k}{ d a_{m} } |\phi_l> 
+  <\phi_l|  \frac{d \phi_k}{ d a_{m} }> 
+< \frac{d \phi_l}{ d a_{m} } |\phi_k> 
-< \frac{d \phi_l}{ d a_{m} } |\phi_k> 
\}
\label{eqn9c}\\
\frac{dE^{SIC}}{d a_{m}}
=\Sigma_{kl}^{\prime} \epsilon^k_{kl}\{ < \frac{d \phi_k}{ d a_{m} } |\phi_l> 
-  < \frac{d \phi_l}{ d a_{m} } |\phi_k>\} \equiv \Sigma_{kl}{'}\epsilon^k_{kl}\Delta_{lk,m},  \label{eqn10}
\end{eqnarray} 
with $\epsilon^k_{kl}=<\phi_l|V_k^{SIC}|\phi_k>$. Due to the fact that the localized orbitals are constrained
to lie in the space of the Kohn-Sham orbitals, this is a generally correct formula and there is no requirement that
Kohn-Sham orbitals need to be self-consistent solutions of a Hamiltonian.
Now to evaluate the derivatives in the above expression it is first necessary to review L{\"o}wdin's  method
of symmetric orthonormalization which proceeds by determining a set of orbitals, referred to here as
intermediate L{\"o}wdin orbitals (ILO) by diagonalizing the overlap matrix of the Fermi orbitals according to:
\begin{eqnarray}
|T_\alpha>=\Sigma_j T_{\alpha j}|F_j> \label{eqn3} \\
\Sigma_{j} S_{ij} T_{\alpha j} = Q_\alpha T_{\alpha i} \label{eqn4} \\
S_{ij}=<F_i|F_j>  \label{eqn5}
\end{eqnarray}
The eigenvalues of the FO-overlap matrix, $Q_\alpha$, tell us how much charge each ILO captures.
From Eqs.~\ref{eqn3},~\ref{eqn4},~\ref{eqn5}, the localized orbitals (LO), designated by {$\phi_k$}, 
are constructed from the ILO and associated eigenvalues according to:
\begin{eqnarray}
|\phi_k>= \Sigma_{\alpha j} \frac{1}{\sqrt{Q_\alpha}}T_{\alpha k}T_{\alpha j} |F_j> \equiv \Sigma_j \phi_{kj}^F |F_j> \label{eqn6}
\end{eqnarray}
As such, 
it follows that:
\begin{eqnarray}
|\frac{ d \phi_k}{d a_{m}}>=|D_{1,km}>+|D_{2,km}>+|D_{3,km}> \equiv \Sigma_{l} {\bf \Delta}_{kl,m} |\phi_l>, \\
|D_{1,km}> = \Sigma_{\alpha j} \frac{1}{\sqrt{Q_\alpha}}T_{\alpha k}T_{\alpha j} |\frac{ d F_j}{d a_m}> = 
\Sigma_{\alpha  } \frac{1}{\sqrt{Q_\alpha}}T_{\alpha k}T_{\alpha m} |\frac{ d F_m}{d a_m}>,  \\
|D_{2,km}> = -\frac{1}{2} \Sigma_{\alpha j} \frac{1}{Q_\alpha^{3/2}}T_{\alpha k}T_{\alpha j}\frac{ d Q_\alpha}{d a_m} | F_j>, \\
|D_{3,km}> = \Sigma_{\alpha j}\frac{1}{Q_\alpha^{1/2}} 
\{
\frac{d T_{\alpha k}}{d a_m} T_{\alpha j} + T_{\alpha k} \frac{d T_{\alpha j}}{d a_m} \} |F_j> . 
\end{eqnarray}
With quite a bit of algebra, which includes a perturbative analysis to determine quantities 
such as 
$d T_{\alpha j}/d a_m $, the quantities $<\phi_l|D_{n,km}>-<\phi_k|D_{n,lm}>\equiv \Delta^n_{lk,m}$ 
are needed to
evaluate Eq.~\ref{eqn10}. The full analysis will be published in a longer paper along with 
a discussion of scaling with system size. Here the final results are given:
\begin{eqnarray}
\Delta^1_{lk,m}=
\Sigma_{\alpha \beta n  } \frac{  
 T_{\alpha k}T_{\alpha m} T_{\beta l}T_{\beta n} 
-T_{\alpha l}T_{\alpha m} T_{\beta k}T_{\beta n} }
{\sqrt{Q_\alpha Q_\beta} } 
\frac{d S_{nm}}{d a_m} \\
\Delta^2_{lk,m}=
<\phi_l|D_{2,km}> - <\phi_k|D_{2,lm}> = 0, \\
\Delta^3_{lk,m}=-\frac{1}{2}
\Sigma_{ \alpha  \beta n } 
\frac{d S_{nm}}{d a_m}\{
T_{\beta n} T_{\alpha m}+
T_{\beta m} T_{\alpha n} \} 
\{T_{\alpha k}T_{\beta  l} 
-T_{\alpha l}T_{\beta  k}\}
\frac{Q_{\beta }^{1/2} -Q_{\alpha}^{1/2}   }
{(Q_\alpha^{1/2}+Q_\beta^{1/2})(Q_\alpha Q_\beta)^{1/2}} 
\end{eqnarray}
with $\frac{d S_{nm}}{d a_m} =  <F_n|\frac{ d F_m}{d a_m}>  \delta_{nm}$. The Kronecker delta arises because the Fermi
Orbital is always normalized~\cite{PRP,FO1,FO3}. Therefore only overlap integrals between different Fermi Orbitals 
change when the FOC 
of one of the Fermi Orbitals changes. The gradient of a Fermi Orbital is in fact a linear combination of the
original Fermi orbitals since the Fermi-Orbital construction always leads to Fermi Orbitals that span the space of the
Kohn-Sham orbitals. However it is also a linear combination of the Kohn-Sham orbitals and since these are orthonormal, the 
derivatives of the overlap integals are most easily calculated by expanding the Fermi orbital derivatives in terms of the 
Kohn-Sham orbitals.
To determine the derivatives of the Fermi-Orbital overlaps in the above expressions it is useful
to use the following:
\begin{eqnarray}
\nabla_{ a_{i \sigma} } F_{i\sigma}({\bf r})=
\frac{\Sigma_{\alpha} \{  \nabla_{a_{i\sigma}} \psi_{\alpha\sigma}
({\bf a}_{i\sigma})\}
\psi_{\alpha\sigma}(\bf r)}
{\sqrt{\rho({\bf a}_{i\sigma}})}
-\frac{ F_{i\sigma}({\bf r}) \nabla_{a_{i\sigma}}\rho(\bf a_{i\sigma}) }
{{2\rho({\bf a}_{i\sigma}})}
\equiv\Sigma_{\alpha}
\{\nabla_{{\bf a}_{i\sigma}} F_{i\alpha}^{\sigma}\}
\psi_{\alpha\sigma}({\bf r})\label{eqn2p}. \\
\nabla_{a_{i\sigma}} F_{i\alpha}^{\sigma}=
F_{i\alpha}^{\sigma} \{
\frac{\nabla_{a_{i\sigma}} 
\psi_{\alpha\sigma}
({\bf a}_{i\sigma})
} { 
\psi_{\alpha\sigma}
({\bf a}_{i\sigma})}
-\frac{ \nabla_{a_{i\sigma}}\rho(\bf a_{i\sigma}) }
{{2\rho({\bf a}_{i\sigma}})}
\}
\end{eqnarray}
The above expression shows that that the gradients of the Kohn-Sham orbitals and spin densities ultimately control
the overlap integrals. As such, even a local expression for the density-functional approximation, such as the one
used here, is "educated" by gradients of the density when the FO-SIC is used.  
With the above equations, the final expression for the derivative of the SIC energy with
respect to a Fermi-Orbital centroid is given by Eq.~\ref{eqn10} with $\Delta_{lk,m}= \Delta^1_{lk,m}+ \Delta^3_{lk,m}$.

\section{Applications to atoms}
                                                                                                                           
For the applications discussed here the PW92 local density approximation is used and the SIC energy 
is constructed according to the Fermi Orbital and the functional form in the PZ paper~\cite{PZ}.
\begin{equation}
E_{xc}^{SIC-PW92} =
-\Sigma_{i,\sigma} \{ U[\rho_{i,\sigma}]+
E_{xc}^{PW92} [\rho_{i,\sigma},0]\}.
\end{equation}
A modified version of NRLMOL~\cite{nrlmol1,nrlmol2,nrlmol3} has been used to perform the calculations. The basis used in these
calculations are those determined in the original gaussian-basis-optimization method of Porezag and 
Pederson\cite{pp} which optimized basis functions for 
the PW92 energy functional.  
These basis sets differ only slightly from the current more widely distributed 
NRLMOL basis sets which were optimized for the PBE-GGA functional. As discussed in Ref~\onlinecite{pp}, 
these basis sets satisfy the
Z$^{10/3}$ theorem, necessary for chemically accurate converged core-level energies, 
and are expected to provide energies that are very close to the converged numerical energies. To obtain Fermi-Orbital
derivatives that were clearly zero at the minimum, the energies were converged to 10$^{-8}$ Hartrees.  In Table II, the total energies
calculated using FSIC are presented and compared to accurate Quantum Monte Carlo results~\cite{Buendia,Caffarel}
and experiment~\cite{Wei}.   The total SIC energy, per spin,
 ranges from -0.5 to -4.0 Hartrees (in Ne and Sr respectively).
In the forthcoming subsections 
a few more details are provided about these 
calculations.
\begin{center}
\begin{table}[htp]
\begin{tabular}{|l|r|r|r|r|r|r|r|}
\hline
Atom     & PW92-LDA& FSIC(PW92)& QMC(2007)& QMC (2014)  &Expt&HOMO$^{FSIC}$& I$_p^{expt}$ \\
\hline
Be       & -14.446  &-14.703   &-14.646  (B) &              & -14.667  &  9.22  &9.32   \\ 
Ne       &-128.230  &-129.268  &-128.892  (B)&         &-128.938  & 24.93  &21.56  \\ 
Mg       &-199.135  &-200.538  &-199.986  (B)&              &-200.054  &  7.62  & 7.64 \\ 
Ar       &-525.939  &-528.522  &-527.391  (B)&         &-527.544  &     17.06  &15.76  \\ 
Ca       &-675.735  &-678.740  &-677.377  (B)&              &          &  5.99  & 6.11  \\
Zn       &-1776.561 &-1782.059 &-1779.119 (B)&-1779.342(S)&         &  9.49  & 9.39 \\ 
Kr       &-2750.133 &-2757.585 &-2753.486 (B)&               &         & 15.11  &14.00 \\ 
Sr       &-3129.437 &-3137.510 &         &               &         &  5.52  & 5.70  \\ 
\hline
\end{tabular}
\caption{Total energies (Hartrees) of atoms for PW92-LDA, FSIC-PW92-LSDA (this work) and QMC (other work). The highest occupied 
eigenvalue is compared to the experimental ionization energy. All calculations in the table correspond to non-relativistic 
Hamiltonians. The QMC results are from Buend\'ia {\em et al} 
and Scemama {\em et al} designated by B and S respectively. The experimental
analysis is due to G. Martin and may be found at Ref.~\onlinecite{Wei}.}
\end{table}
\end{center}

\subsection{Neon and Argon}

Neon and Argon represent relatively simple cases since $sp^3$ hybrids have already been shown to minimize the SIC
energy without breaking the three-fold degeneracy of the highest-occupied orbitals. For Ne the $1s$-centroid was placed
at the origin and the $2sp$-centroids were placed at the vertices of a tetrahedron. For Ne, the distance from the origin
to the vertices of the tetrahedron was found to be 1.053 Bohrs. At this position the derivative on each Fermi-Orbital was smaller
than 0.0022eV/\AA.  The $2p$ eigenvalues were found to be -24.9 eV which is in reasonable but not excellent agreement 
with experimental ionization potential.  
The total energy of -129.268 Hartree is much closer to the nonrelativistic QMC results (-128.992 to -128.938)
of Buend\'ia and experiment~\cite{Wei} respectively.  

For Argon, the optimized tetrahedral vertices of the $2sp$ Fermi-Orbitals were found to be at 0.391 Bohr. The $3sp$ 
Fermi-Orbital tetrahedral vertices were found to be at 1.345 Bohr. For Ar, at convergence the largest 
Fermi Orbital derivative was found to be:
0.0007 eV/\AA~ which is much smaller than necessary for convergence of the total energy.  
The total energy of -528.522 Hartree is again close to the QMC resul522nd significantly improved
over the LDA energies.  The 3p eigenvalues (-17.06 eV) were found to be in good agreement with the experimental ionization energy
of 15.76 eV.  The results discussed in this section are in qualitative accord with 
the earlier exchange-only results of Pederson, Heaton and Lin~\cite{PedJCP88}. 
However a the inclusion of a correlated functional leads to total energies that are even 
lower than experimental values.

\begin{figure}[htp]\centering{ 
\includegraphics[scale=0.60]{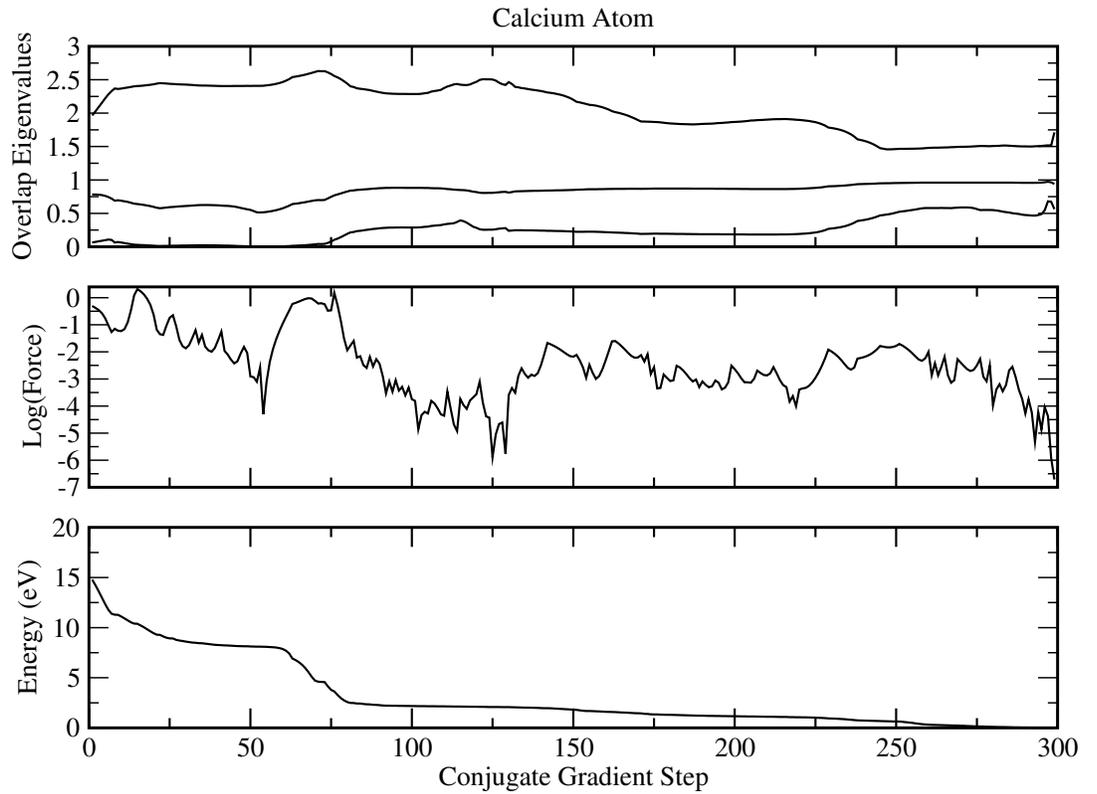} }
\caption{Three panels show the difference of the SIC energy relative to the minimum (lower), 
the natural log of the norm of Fermi-Orbital gradients (middle), and information about the
the lowest and  highest ILO eigenvalues and geometric mean of the ILO eigenvalues is (upper). The geometric mean would be 
exactly unity if FOC are found that that lead to orthonormal Fermi Orbitals. Initial optimization (steps 0 to 125) constrained FOC's to have trigonal 
symmetry. This constraint was then completely relaxed (steps 125-300). Total SIC-LSD and LSD energies are compared to QMC results in Table 1.  }
\end{figure}
\subsection{Beryllium, Magnesium and Calcium}                    
Very briefly it is useful to test the FO-method on Be since it is the simplest case for a 2-electron (per spin) system and 
since it is actually a challenging case for DFT. 
The Be atom is known to be highly susceptible to treatment of correlation due to the
near degeneracy of the occupied $2s$ and unoccupied $2p$ electrons that occurs experimentally.
Indeed, on a percentage basis, deviation in total energy between SIC-LSD and the QMC/experimental
results is larger than for all the atoms considered here.
The energies in
Table 1 show that a total energy of -14.703 and a $2s$ eigenvalue of 9.22 eV which is in very good agreement with the
ionization energy. 
For the lowest-energy configuration with vanishing Fermi-Orbital derivatives it is verified, as expected,  
that the localization
equations are not satisfied. This verifies that the Fermi-orbital-based formalism for SIC, with the constraint that the energy is
explicity invariant under unitary transformations, does not reduce to the standard formulation in the simplest possible limit.

This column of the periodic table presents the first case where there is a degree of geometical 
frustration in determining physically
appealing Fermi-Orbital centroids. In this section results for Mg and Ca are discussed but Sr is discussed in the following
section as it contains $3d$ electrons.  The three valences of $s$ electrons and two valences of $p$ electrons do not allow for simple
$sp^3$ hybridization in this column of the periodic table. 

For Mg it is found that the FOCs which minimize the energy correspond to positions that may be viewed as a tetrahedron of
$3s3p^3$ hybrids with an additional FO capping one of the triangular faces.  A slightly less stable solution composed of a
$3s2p_z$ hybrids, a trio of $2sp_{xy}$ hybrids and an $s$-like Fermi-orbital was also found.  The HOMO eigenvalue associated
with the lowest energy solution (7.62 eV) was in much better agreement with experiment (7.64 eV) than the higher energy 
solution (9.52 eV).  The derivatives on the FOC were again converged to a very small value (~0.005 eV/\AA). 

For the calcium atom  additional detail is provided
since it has a bit more complexity than some of the lighter atoms
but no $3d$ electrons. As a relatively good test of the Fermi-Orbital derivatives, 
the original guess for the FOCs was that the localized
orbitals hybridized as a $1s$ core state, a  pair of $3sp_z$ orbitals, 
a trio of $2sp_{xy}$ orbitals, and a quartet of $4s-3p^3$ orbitals. 
Such starting points are easily accomplished by constructing sets of initial Fermi-Orbital 
centroids that coincide with (1) the origin, (2) a small 
bicapped triangular platonic solid, (3) a larger tetrahedron respectively.   In Figure 1, the SIC energy 
(relative to the minimum) and the logarithm of the norm of the Fermi orbital gradient are shown as a function of 
conjugate gradient step. The figure shows that the derivatives vanish at the lowest energy.  

\subsection{Zinc, Krypton and Strontium}

The author's primary motivation for returning to the self-interaction correction was that large systems containing mixtures of
transition-metal ions, nearly-free-electron metal atoms, and ligands composed of first-row atoms are challenging but
not insurmountable  within the 
framework of density-functional theory~\cite{molmag,canali,park,cheng}. These challenges stem from the
atom-dependent mismatches of the chemical potential which make systematic calculations on such devices/systems
difficult.  Such mismatches lead to incorrect $s-d$ occupations in atoms,
incorrect spin-ordering in nickel-containing molecular magnets~\cite{park,cheng} 
and spurious charge transfer between metallic leads
and molecular islands in molecular circuits.~\cite{canali} 
Transition-metal containing systems, and their interactions with 
molecules composed of light atoms are also of interest to understanding catalytic processes and separation of weakly
interacting gases. Such open-metal sites may be challenging since their entirely  
open-shell counterparts (isolated transition metal atoms) are 
challenging~\cite{janak,hoyer}.  
In this section some preliminary results on closed shell-atoms containing $d$ electrons are 
presented. The Fermi-orbital centroids are available upon request and will be broadcast, along with a simple 
executable version of NRLMOL-based SIC program from a website soon. Given that in one case (Mg), two low-energy
solutions were found,  this will allow others to look for additional 
solutions for these and other atoms and to help to determine whether the solutions here are indeed the absolute minima 
within the FO-formulation of SIC.

The total energies of 
these atoms are presented in Table 1 and are also found to be in good agreement with  QMC results. The HOMO eigenvalues
seem to be in excellent agreement with measured ionization energies.  For very bad guesses of
Fermi orbital positions the eigenvalues have absolutely no similarity to the converged results which reproduce, to one
percent, the 3-fold and 5-fold degeneracies, expected in closed-shell atoms. It remains to be determined if, as
in the case of Ne, Ar, Mg, and Ca, these degeneracies can be fully restored by the Fermi-Orbital with the best Fermi-Orbital
centroids. Results from averaged SIC calculations (e.g. a slightly different functional form) suggest that such a modification
could be possible~\cite{MaiDinh,Suraud}.

\section{Summary}
Expressions for the derivatives of the self-interaction-corrected density-functional energy have been derived and
presented. The analytic expressions are then numerically validated by showing, in conjunction with the Perdew-Zunger
formulation of self-interaction corrected DFT, that the derivatives vanish when the energy is minimized and that the derivatives
may be used to navigate to the minimum energy.
The work here demonstrates proof of principle and, through comparison to accurate QMC results, 
finds that the total energies of self-interaction corrected LDA results are improved. Based on past recollections~\cite{PedJCP88},
and recent discussions~\cite{Susi}, it can not yet be said that the FO-based construct will be faster
than the original localization-equation-based method~\cite{PedJCP84,PedJCP88}. In comparison to the real cases, the solutions are
energetically very close but not identical.
Consideration of the improved atomic total energies, demonstrated here, and improvements in 
molecular cohesive energies, demonstrated in Ref.~\onlinecite{PRP}, give reason for a degree of optimism. However further
systematization would be desireable prior to actually optimizing a new asymptotically correct 
functional that is explicitly self-interaction corrected using the FO-based approach.

While the work presented here can all be reproduced (including optimizations) in a day on 
a laptop computer, the LDA calculations are much faster than this. It is clear 
that off-the-shelf conjugate-gradient methods, with initial step sizes tuned for optimization of molecular geometries, may  
need to be preconditioned for this type of problem and that additional work will be 
needed to determine when the Fermi-orbital derivatives effectively vanish (rather than the strong convergence criteria
used here).
Futher, additional work aimed at determining transferable starting points will also
need to be pursued for larger-scale applications.  
However, additional analysis of the functional form of the SIC energy must be perfomed in parallel since 
the numerics could become significantly easier if "softer" or nearly nodeless SIC functionals could be developed.

In regard to symmetry breakings, for closed-shell atoms devoid of $d$ electrons, all orbital 
degeneracies expected from general symmetry arguments are restored by both forms of SIC. 
However, as for the case of the now standard SIC approach, The FO-SIC  methodology does not yet lead to perfect five-fold
degeneracies for closed-shell atomic systems.
Even in the absence of hybridization between states of different angular momentum, 
it is difficult to have an orbital-by-orbital SIC correction that automatically restores the degeneracies to 
closed-shell $d-$ and $f-$ electron systems. 
Arguments 
could be made that degeneracy breaking may not be an issue since it is only the highest
occupied eigenvalue that has physical meaning within DFT. 
On the positive side, for atoms containing $d$-electrons, there are clear indications that the 
self-interaction correction for the $d$ electrons are much larger than for the outer s-electron. This feature is needed
to account for some of the problems that have been observed in DFT-based calculations on systems where there is 
competition in $3d-4s$ shell fillings.~\cite{park,cheng,hoyer}  
Such open-shell systems will provide good benchmark challenges for SIC methods.

The FOC-methodology, which always delivers real orbitals in closed-shell systems, leads to a conundrum since
there are a variety of results, including recent applications to benzene~\cite{Iceland}, that suggest that smoother localized
densities, such as those obtained from complex orbitals, lead to better agreement with experiment. 
As such, the Fermi-Orbital filtering which disallows complex orbitals provides a double-edge sword. Disallowing complex
Bloch functions guarantees size-extensivity in solids composed of well-separated atoms in cases where the SIC energy 
of an atom turns positive.  Further in simple closed-shell atoms, 
the Fermi-orbital-based methodology will not allow for symmetry breaking of the p-orbitals for any functional since it does
not allow for consideration of orbitals constructed from spherical harmonics (which lead to different SIC shifts for the $m_s=0$ and
$m_s=\pm 1$ states).  However, since either the localization transformation or Fermi-orbital-based constraint must be chosen
for any energy functional that exhibits any degree of orbital dependence, it is more likely that these relatively small disparities will
be overcome by reconsidering the construction of the SIC energy {\em per se}  and that the determination of whether the 
FO-based approach should replace the localization equations will be based upon other considerations. From the perspective of
unitary invariance as a strong constraint,
there seems to be a strong reason for choosing the Fermi orbital-based formulation. But on the other hand, some of the
most accurate multi-configurational methods essentially proceed by accepting that unitary invariance is not the most important
symmetry. This might argue for a 
formulation more similar to the standard perspective.
In fact, the original derivation and uses of unified Hamiltonians~\cite{HeaPRB83,Iceland}, commonly used for SIC calculations 
since 1983, stem from even earlier use in the then nascent field of MCSCF (See Ref.~\onlinecite{HarJPB83,HeaPRB83} and references therein).
Reasons for searching for new forms of the SIC functional will be discussed in an upcoming community-generated
paper which carefully examines a large amount of numerical results showing promise for cases where 
the SIC functional is constructed from orbital densities that are in close energetic and spatial proximity to one another. 
The FO construction allows for the straightforward means for
defining local orbital energies and determining their spatial and energetic proximity. 

The average of two real localized orbital densities
is of course exactly equal to the density constructed from a complex orbital determined from the same
two orbitals.
Perhaps a new prescription that provides 
a straightforward well-defined means for creating average nearly nodeless localized densities will effectively allow 
for the complex-localized orbitals densities, suggested by J\'onsson~\cite{simon1,simon2,Iceland} 
and K{\"u}mmel~\cite{hoffmann,Kummel}, and begin to address the critique provided by K{\"u}mmel that suggests
that the SIC functional itself should never have densities composed of orbitals with nodes. The FO-construction provides
new ways for automating the determination of localized densities so it may be that other SIC functionals which satisfy the 
constraints suggested in Ref.~\onlinecite{RLT} are possible.                
Thinking about such an averaging procedure would not
be inconsistent with this version of SIC or the two-step standard version of SIC since it would only change the definition 
of the off-diagonal Lagrange multipliers in Eq.~\ref{eqn10}. However, it is 
likely that the derivatives themselves would become smoother
and furnish Fermi-Orbitals and SIC energies that were slightly more ambivalent to their positions. 

Finally, it may also be interesting to seek numerical proof that a class of 
"most physical" solutions are attainable within both formulations. For example if it is possible to consider only FO's that lead to
an othonormal overlap matrix, (e.g. values of unity for all ILO eigenvalues), 
one would then have a set of solutions that are possible within both formulations and these solutions 
would almost definitely coincide with the most loved localized orthonormal orbitals in physics and chemistry.  
Further inquiries along these directions may 
very well allow for a step in the direction of merging the "road less traveled" with the "road more traveled" 
discussed in Ref.~\onlinecite{RLT}.   But a significant amount of analytical work and analysis of computational results 
is required to address these points.
                                                                     
\section{Acknowledgements}
MRP thanks Dr. S. Lehtola for providing unpublished data, for discussions, and for a careful reading of 
the manuscript. MRP thanks J. Sun, K. Bowen, C.C. Lin, J.P. Perdew, A. Ruzsinszky, and S. K{\"u}mmel  
for interest in this work and for comments and critique.


\newpage                                                                               
\begin{references}
\bibitem{PRP}M.R. Pederson, A. Ruzsinszky and J.P. Perdew, J. Chem. Phys. {\bf 140  } 121105 (2014). 
\bibitem{Low}P.O. L{\"o}wdin, Rev. Mod. Phys. {\bf 34}, 520 (1962).
\bibitem{FO1} W.L. Luken and D.N. Beratan, 
Theor. Chim. Acta, {\bf 61}, 265-281 (1982).
\bibitem{FO3} W.L. Luken and J.C. Culberson, Theor. Chim.  Acta, {\bf 66} 279-283 (1984).
\bibitem{PW92}J. P. Perdew, J. A. Chevary, S. H. Vosko, K. A. Jackson, M. R. Pederson, D. J. Singh, and C. Fiolhais,
Phys. Rev. B {\bf 46}, 6671 (1992).
\bibitem{PZ} J.P. Perdew and  A. Zunger, Phys. Rev. B {\bf 23}, 5048 (1981).
\bibitem{HK1} P. Hohenberg and W. Kohn, Phys. Rev. {\bf 136}, B864 (1964).
\bibitem{HK2} W. Kohn and L.J. Sham, Phys. Rev. {\bf 140}, A1133 (1965).
\bibitem{RE63}C. Edmiston and K. Ruedenberg, Rev. Mod. Phys. {\bf 35}, 457 (1963).
\bibitem{HarJPB83}J.G. Harrison, R.A. Heaton,  and C.C. Lin, J. Phys. B {\bf 16}, 2079 (1983).
\bibitem{HeaPRB83}R.A. Heaton, J.G. Harrison and C.C. Lin, Phys. Rev. B {\bf 28}, 5992 (1983).
\bibitem{PedJCP84} M.R. Pederson, R.A. Heaton, and C.C. Lin, J. Chem. Phys. {\bf 80}, 1972 (1984).
\bibitem{PedJCP85} M.R. Pederson, R.A. Heaton, and C.C. Lin, J. Chem. Phys. {\bf 82}, 2688 (1985).
\bibitem{PedJCP88}M.R. Pederson and C.C. Lin, J. Chem. Phys. {\bf 88}, 1807 (1988).
\bibitem{temmerman}A. Svane, V. Kanchana, G. Vaitheeswaran, G. Santi, W.M. Temmerman, Z. Szotek, P. Strange,
 and L. Petit, Phys. Rev. B {\bf 69}, 054427 (2004).
\bibitem{simon1}S. Kl{\"u}pfel, P. Kl{\"u}pfel and H. J\'onsson, J. Chem. Phys. {\bf 137}, 124102 (2012).
\bibitem{simon2} S. Kl{\"u}pfel, P. Kl{\"u}pfel, and H. J\'onsson, Phys. Rev. A
{\bf 84}, 050501(R) (2011). 
\bibitem{hoffmann}D. Hoffmann and S. K{\"u}mmel, J. Chem. Phys. {\bf 137}, 064117 (2012).
\bibitem{MaiDinh}P.M Dinh, C.Z. Zhao, P. Kl{\"u}pfel, P.G. Reinhard, E. Suraud, M. Vincedon, J. Wang, and F.S. Zheng, 
Eur. Phys. Jour. D {\bf 68} 239 (2014).
\bibitem{Suraud}J. Messud, P.M. Dinh, P.-G. Reinhard and E. Suraud, Phys. Rev. A {\bf 80}, 044503 (2009).
\bibitem{RLT}M.R. Pederson and J.P. Perdew, ``Self-Interaction Correction in Density
Functional Theory: The Road Less Traveled", $\Psi_K$ Newsletter Scientific Highlight
of the Month, February (2012) (See: http://www.psi-k.org/newsletters/News\_109/Highlight\_109.pdf)
\bibitem{Kummel}D. Hofmann, S. Kl{\"u}pfel, P. Kl{\"u}pfel and S. K{\"ummel}, Phys. Rev. A {\bf 85} 065214 (2012).
\bibitem{Iceland} S. Lehtola and H. J\'onsson, J. Chem. Theory Comput., Article ASAP
DOI: 10.5324 (2014).    
\bibitem{Dabo}G. Borghi, A. Ferretti, N.L. Nguyen, I. Dabo and N. Marzari, Phys. Rev. B {\bf 90}, 075135 (2014).
\bibitem{nrlmol1} M. R. Pederson and K. A. Jackson, Phys. Rev. B, {\bf 41}, 7453 (1990). 
\bibitem{nrlmol2} K. A. Jackson and M. R. Pederson, Phys. Rev. B {\bf 42}, 3276 (1990). 
\bibitem{nrlmol3} M.R.  Pederson, D.V. Porezag, J. Kortus and D.C. Patton, Phys. Stat. Solidi B {\bf 217},  197 (2000).
\bibitem{pp}D. Porezag and M.R. Pederson, Phys. Rev. A {\bf 60} 2840 (1999).
\bibitem{Buendia} E. Buend\'ia, F.J. G\'alvez, A. Sarsa, Chem. Phys. Lett. {\bf 428} 241 (2006).
\bibitem{Caffarel} A. Scemama, T. Applencourt, E. Giner and M. Caffarel (arXiv:1409371v1 [phys.chem-ph] (12 September 2014).
\bibitem{Wei} Exact non-relativistic atomic energies, determined from successive ionizations, and relativistic corrections  
may be found at: $http://www.weizmann.ac.il/Organic\_Chemistry/martin/atoms.shtml$. Similar experimental
energies, from Ref. [23], of  Ref.~\onlinecite{PedJCP88} may be found.
\bibitem{molmag}A. Postnikov, J. Kortus, and M.R. Pederson, Phys. Stat. Solidi (b) {\bf 243}, 2533 (2006).
\bibitem{canali}J.F. Nossa, M.F. Islam, C.M. Canali, M.R. Pederson, Phys. Rev. B {\bf 88}, 224423 (2013).
\bibitem{park}K. Park, E.-C. Yang, and D.N. Hendrickson, J. Appl.  Phys. {\bf 97}, 10M522 (2005).
\bibitem{cheng}C. Cao, S. Hill, and H.P. Cheng, Phys. Rev. Lett. {\bf 100}, 167206-4 (2008).
\bibitem{janak}J.F. Janak, Phys. Rev. B {\bf 18}, 7165 (1978).
\bibitem{Susi} S. Lehtola, Unpublished results using the basis sets of Ref.~\cite{pp}.
\bibitem{hoyer} C.E. Hoyer, G. Li Manni, D. Truhlar and L. Gagliardi, J. Chem. Phys. {\bf 141} 204309 (2014).
\end{references}
\end{document}